\documentclass[12pt]{article}

\usepackage{graphicx} 

\begin{document}
\centerline{\bf Unusual ferromagnetism in Ising and Potts model}

\centerline{\bf on semi-directed Barab\'asi-Albert networks}

\bigskip

\centerline{ Muneer A. Sumour$^{1}$ and F. W. S. Lima$^{2}$ }

\bigskip

\noindent
$^{1}$Physics Department, Al-Aqsa University, P.O.4051, Gaza, Gaza Strip, Palestinian Authority, \\
$^{2}$Dietrich Stauffer Computational Physics Lab, Departamento de F\'{\i}sica, \\
Universidade Federal do Piau\'{\i}, 64049-550, Teresina - PI, Brazil.\\

\bigskip

e-mail: msumoor@alaqsa.edu.ps, fwslima@gmail.com 

\bigskip

{\small Abstract: We check the existence of a spontaneous magnetisation of 
Ising and Potts spins on semi-directed Barabasi-Albert networks by Monte Carlo 
simulations. We verified that the magnetisation for different temperatures $T$ 
decays after a characteristic time
$\tau(T)$, which we extrapolate to diverge at positive
temperatures $T_c(N)$ by a Vogel-Fulcher law, with $T_c(N)$ increasing
logarithmically with network size $N$.}

\bigskip


\bigskip

\section{Introduction}

The Ising and Potts model
has been used during a long 
time as a "toy model" to test and to improve new algorithms and 
methods of high precision for calculation of critical exponents in 
Equilibrium Statistical Mechanics  using the Monte Carlo method
as Metropolis \cite{me}, Swendsen-Wang \cite{s-w}, 
Wang-Landau \cite{w-l} algorithms. The Ising model was
already applied to scale free networks \cite{BA} or 
{\it undirected} Barab\'asi-Albert networks (UBA), 
where simulations \cite{A_H_S}
indicate a Curie temperature increasing logarithmically 
with increasing system size $N$. Different from \cite{A_H_S}, Sumour 
{\it et al.} \cite{sumuor1,sumuor2} studied the Ising model on
a {\it directed} Barab\'asi-Albert network (DBA) 
using standard Glauber kinetic Ising models
 on fixed networks. They confirmed the 
asymptoptic Arrhenius extrapolation $1/\ln\tau$ $\propto$ $T$
for the time $\tau$ until the first sign change of the magnetisation,
meaning that at all finite temperatures 
the magnetisation eventually vanishes, i.e., no ferromagnetism.

In the present work, we study the critical behaviour
of Ising and Potts model on {\it semi-directed} Barab\'asi-Albert
network (SDBA) recently studied by Sumour and Radwan \cite{sumuor3},
where now the number $N(k)$ of nodes with $k$ links each decays as
$1/k^{\gamma}$ and the exponent $\gamma$ decreases from $3$ to $2$
for increasing $m$ where $m$ is the number of old nodes  which a new node
added to the network selects to be connected with. This behaviour is totally 
different from $UBA$ and $DBA$ scale free networks where  $\gamma=3$ is
universal, i.e., independent of $m$. For both Ising and Potts
model in our results no usual phase transition 
was found, similar to \cite{A_H_S,sumuor1,sumuor2}.
\section{Semi-Directed Barab\'asi-Albert networks}

Both UBA and DBA networks are grown such that the 
probability, of a new site to 
be connected to one of the already existing sites, 
is proportional to the 
number of previous connections to this already 
existing site: the rich get 
richer. In this way, each new site selects exactly 
$m$ old sites as neighbours.
In a UBA network \cite{A_H_S}, the neighbour 
relations for the spin interactions were such that if A has 
B as a neighbour, B has A as a neighbour, while 
for DBA in general B then does
not have A as a neighbour.

In the DBA and UBA network \cite{sumuor1,sumuor2}, if
a new node selected $m$ old nodes as neighbours,
then the $m$ old nodes are added to the Kert\'esz list \cite{stauffer},
and the new node is also added $m$ times to that list.
Connections are made with $m$ randomly selected 
elements
of that list. If one would only add the old 
nodes to the list,
then only the initial core can be selected as 
neighbours, which is not
interesting. But if one adds to the list the m old nodes 
plus once the new node, one has a semi-directed network: SDBA. 
(For usual BA networks, the new node is added $m$ 
times to the Kert\'esz list.)
\section{Model and simulation }

\subsection {Two versions SDBA1 and SDBA2}

Our first version SDBA1 builds the network in 
the way of \cite{sumuor3}. The new
node $n$ selects $m$ sites $j$ which $n$ will 
all influence, while $n$ will be 
influenced only by the first selected $j$. 
Our second version
SDBA2 inverts the direction of the spin interaction: 
The new node $n$ selects $m$ 
sites $j$ which will all influence $n$, while 
$n$ will influence only the first
selected $j$.

We simulate networks with $N$ nodes $i$, with
 spins $S_i$ on each node.
For both Ising and Potts model on SDBA the evolution
in time is given by single spin-flip
Glauber dynamics with a probability $p$ given by
\begin{equation}
p = 1/[1+\exp(2\Delta E/k_BT)], 
\end{equation}
with energy change $\Delta E$ to be defined below 
through eq. (2) and (3). 
Here, the time is defined as one Monte Carlo step (MCS), where
 one MCS is accomplished after all $N$ spins are updated; and
we denote the final Monte Carlo step number as MCSN.
(We use the same number of iterations for equilibration.)
The error bars are usually smaller than the size of symbols,
so we cannot put them into the figures.
The statistical errors were evaluated from $10$ to $100$ samples of initial
configurations and with $4,000$ to $100,000$ Monte Carlo steps ( thermal error).
\subsection{Ising model on SDBA networks}

The Ising interaction energy is given by
\begin{equation}
E=-J\sum_i\sum_j S_iS_j,
\end{equation}
where $S_{i}=\pm 1$ and the inner sum (also in eq. (3)) runs over 
all neighbours $j$ of node $i$.
The magnetisation defined for this model is
$\sum_{i=1}^{N} S_i/N$. 
\subsection{Potts model on SDBA networks}

For Potts model the interaction energy is
\begin{equation}
E=-J\sum_i\sum_j\delta_{S_{i}S_{j}},
\end{equation}
with Kronecker's delta and $S_i = 1,2, \dots q$. Again,
to study the critical behaviour we define 
the magnetisation 
as $(qM-N)/((q-1)N)$, where $M$ is the 
largest of the $q$ numbers
of spins $S_i$ in one of the $q$ directions 
$1,2,\dots,q$, at each iteration. 
\bigskip
\begin{figure}[!htb]
\begin{center}
\includegraphics[angle=0,scale=0.46]{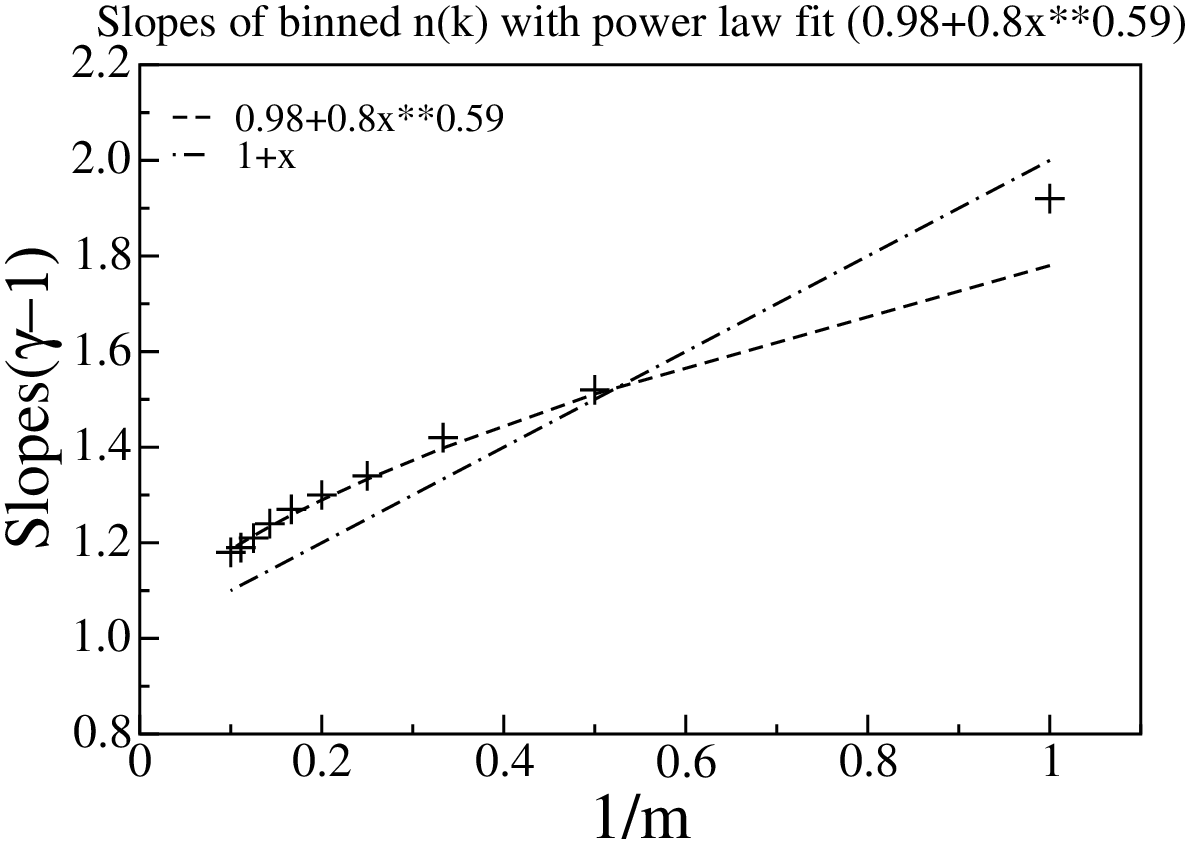}
\end{center}
\caption{Plot of $\gamma(m)-1$ versus $1/m$ with power-law :
$\gamma -1 = 0.98+0.8/m^{0.59}$; $N=400,100,000$ nodes.}
\label{Fig0}
\end{figure}
\bigskip
\section{ Results and discussion}

\subsection{Ising model}

We use the FORTRAN program as in our appendix (1), with different 
$m$. The number of nodes $N$ added to the initial
core of $m$ nodes is 10000 to 50000, and MCSN = 100,000 
iterations were made. First we measure the number $N(k)$ of nodes
influenced by $k$ neighbours in SDBA2, analogous to \cite{sumuor3}
for SDBA1. In Fig. \ref{Fig0}
we plot for each $m$ value (including $m=1$, not shown) we plotted 
double-logarithmically the observed numbers of nodes with at least 
$k$ links each and determined the decay exponents by the 
slopes $\gamma(m)-1$ versus $1/m$,
 which makes clearer the possible extrapolation towards 
infinite $m \; (m=\infty, 1/m=0)$. Maybe the true 
exponents $\gamma(m)$ equal $2 + 1/m$ since $m=1$ 
should give the standard (undirected) exponent 
$\gamma=3$. The deviations from this formula 
(straight line in Fig. \ref{Fig0}) are not much larger than our 
systematic errors. As an alternative to the 
linear behaviour also a power-law fit to $m>1$ is shown. 
We see that the new power-law fit agrees very 
nicely with the data except for the 
standard BA model (undirected) case $m=1$. The behaviour 
of the exponents for much larger $m$ is discussed elsewhere
\cite{preprint} and differs
appreciably between SDBA1 and SDBA2.
\begin{figure}[!ht]
\begin{center}
\includegraphics[angle=0,scale=0.48]{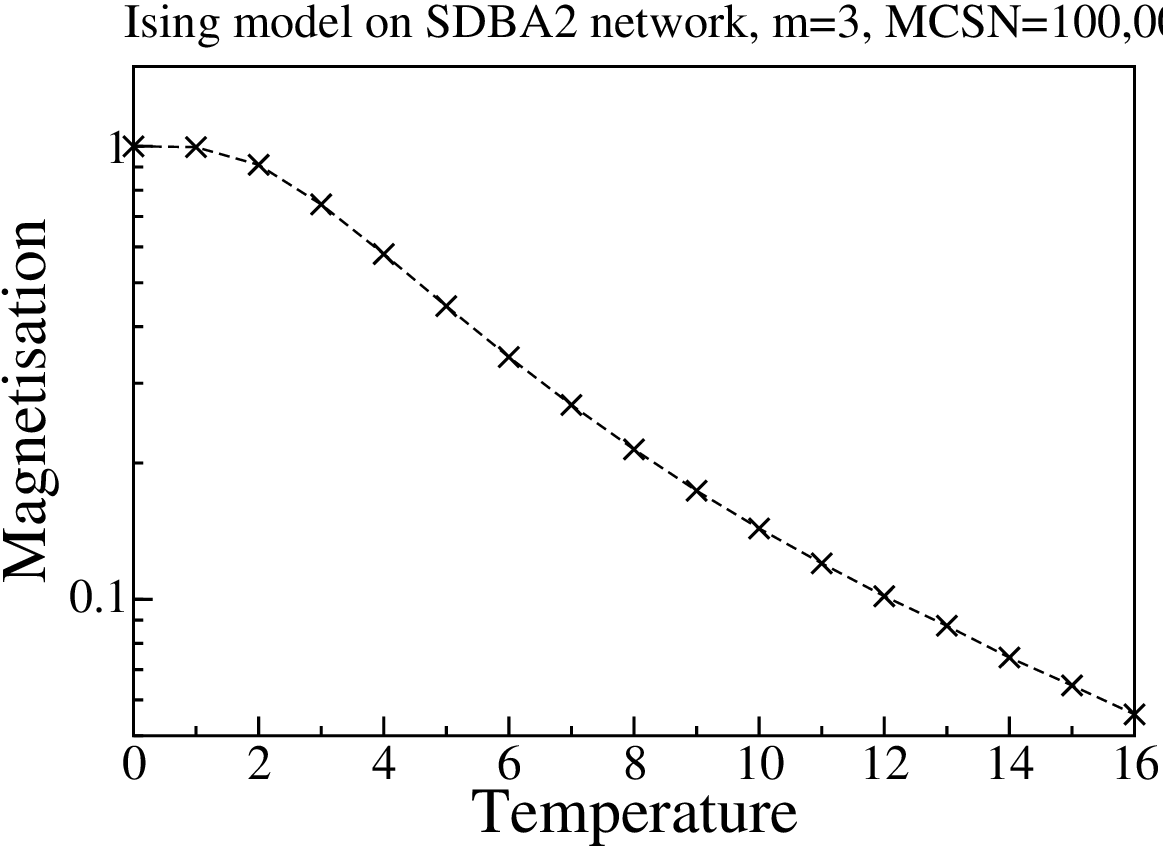}
\end{center}
\caption{Semi-logarithmic plot of magnetisation versus $T$ for $m=3$ and $4000$ nodes}
\label{Fig1}
\end{figure}
\bigskip
\begin{figure}[!hb]
\begin{center}
\includegraphics[angle=0,scale=0.48]{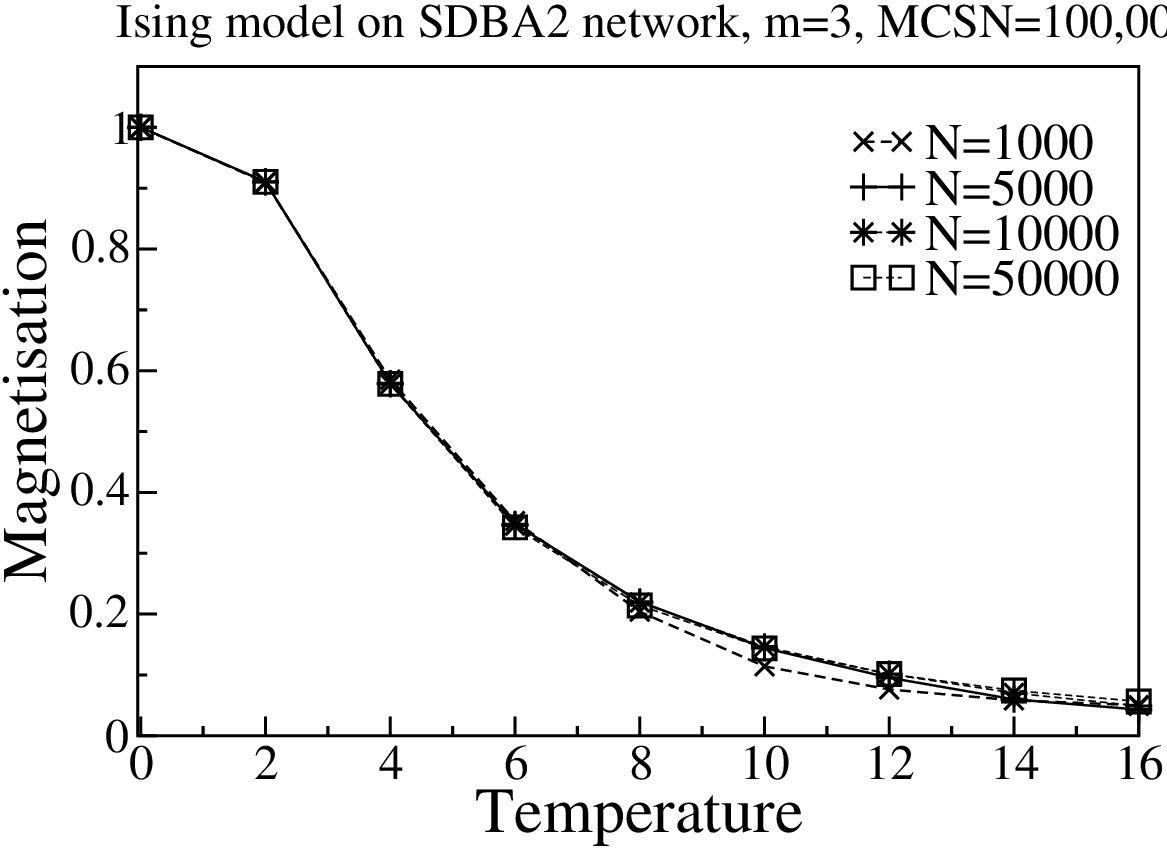}
\end{center}
\caption{Plot of the magnetisation versus $T$ for $m=3$ and
different system sizes $N = 1000$ to $50,000$, $MCSN = 100,000$.}
\label{Fig3}
\end{figure}
\bigskip
\begin{figure}[!hbt]
\begin{center}
\includegraphics[angle=0,scale=0.58]{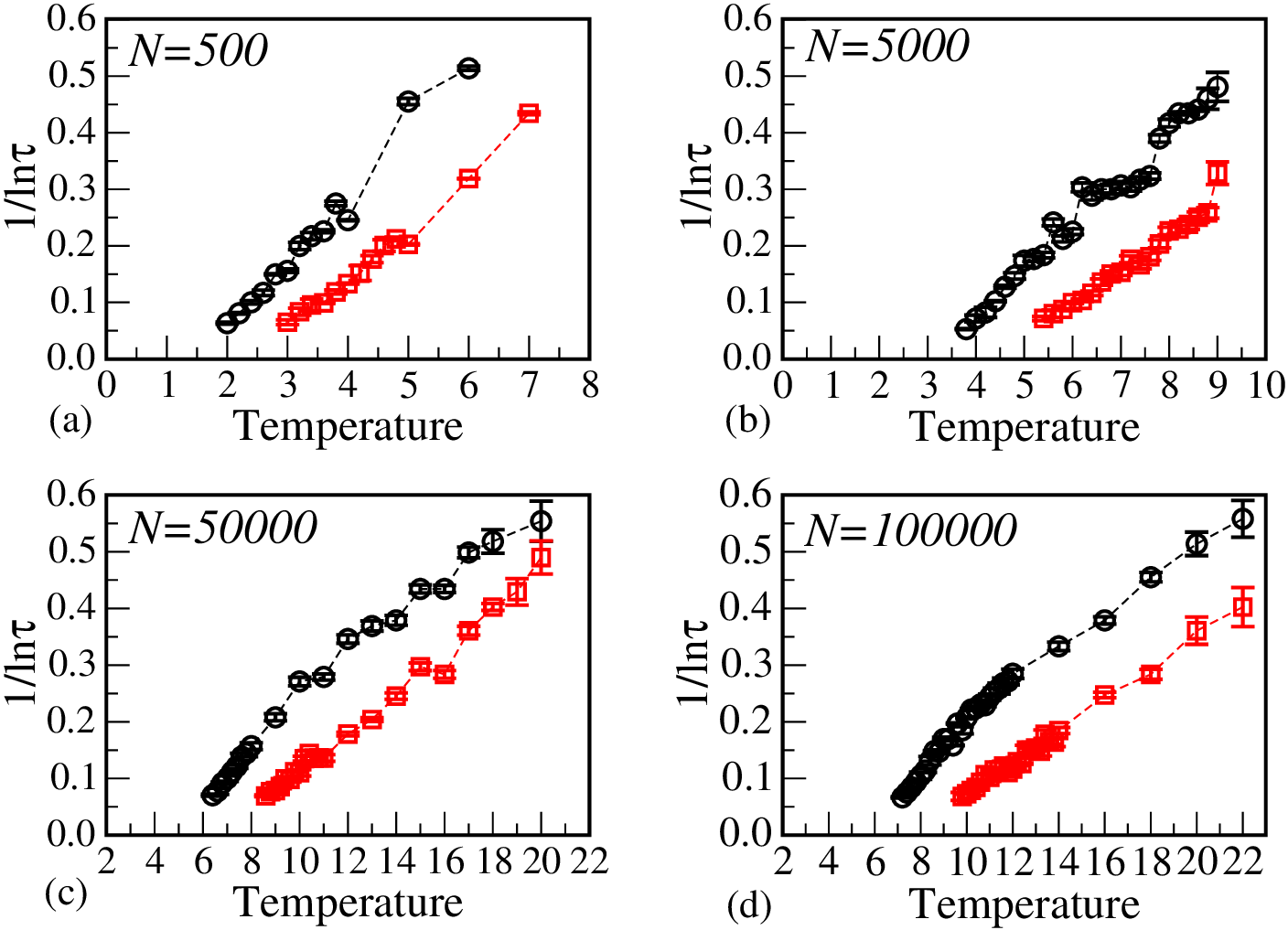}
\end{center}
\caption{Reciprocal logarithm of the relaxation times 
versus temperature for
SDBA1(circle) and SDBA2 (square) networks and Potts model 
with $q=2$ (Ising), $m=2$ initial neighbours and 
$N=500$ (a), $5000$ (b), $50000$ (c), and $100000$ (d) sites.}
\label{Fig10}
\end{figure}
\bigskip
\begin{figure}[!ht]
\begin{center}
\includegraphics[angle=0,scale=0.46]{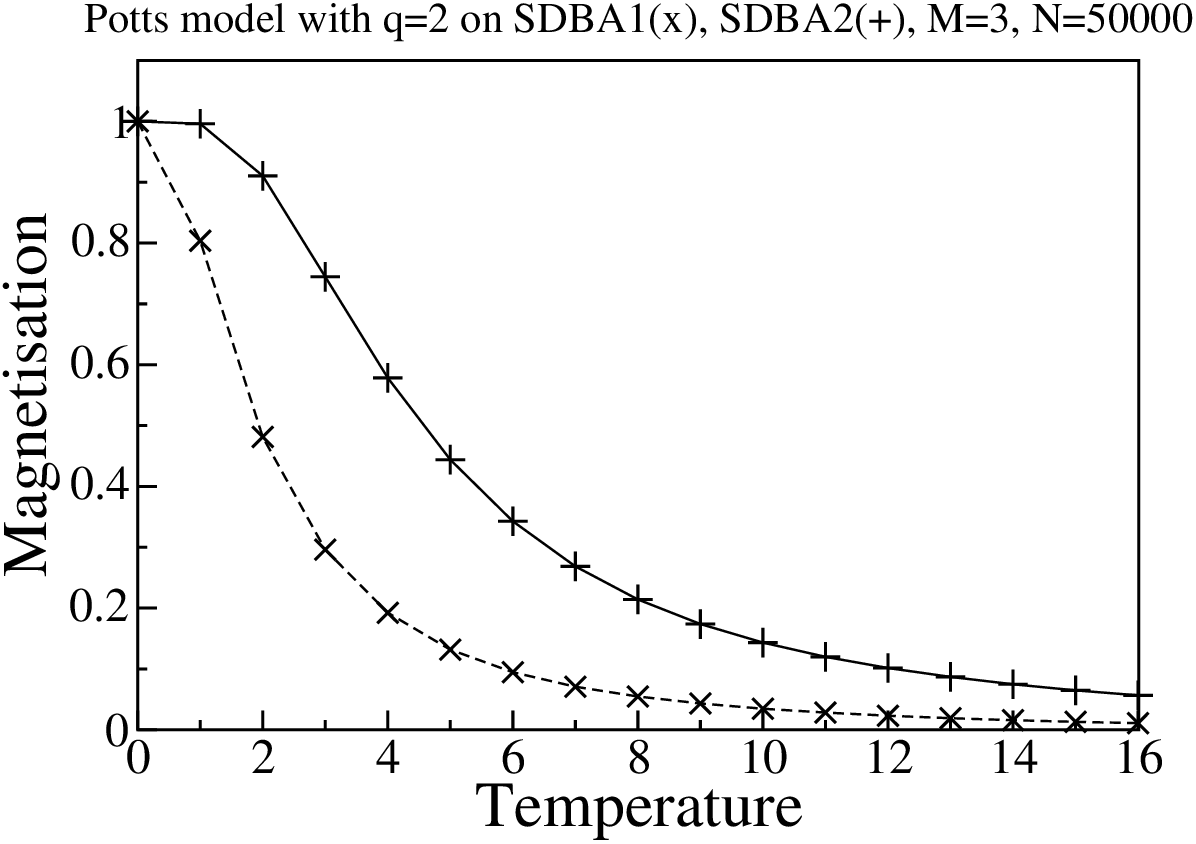}
\end{center}
\caption{Plot of the magnetisation versus temperature  on 
SDBA1($\times$) and SDBA2(+) network for Potts model with $q=2$ states,
$m=3$ initial neighbours,
 $N=50000$ sites and $MCSN = 100,000$ iterations.}
\label{Fig40}
\end{figure}
\begin{figure}[!hb]
\begin{center}
\includegraphics[angle=0,scale=0.46]{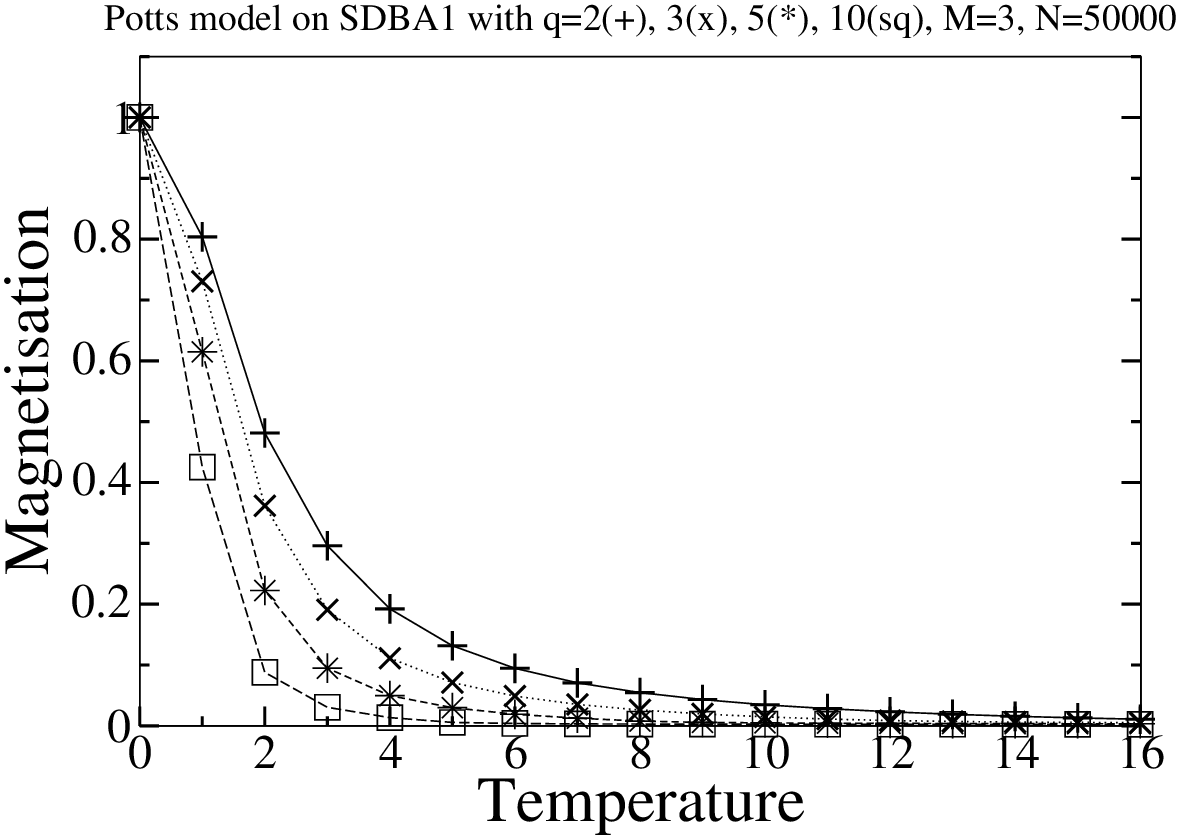}
\end{center}
\caption{Plot of the magnetisation versus $T$ 
for different values of 
$q = 2(+),3(x),5(*)$, and 
$10$(square) on SDBA1 network for $N=50000$ 
sites and $m=3$.}
\label{Fig50}
\end{figure}
Fig. \ref{Fig1} shows the magnetisation as a function of 
temperature $(T=0,1,2,\dots, 16)$. The roughly exponential decay is similar
to \cite{A_H_S}.
Then we change the initial number of neighbours $m=1,3,5,7$ 
with system size $50000$ and 
$MCSN = 4,000$ to $100,000$ iterations. 
\bigskip
\begin{figure}[!ht]
\begin{center}
\includegraphics[angle=0,scale=0.46]{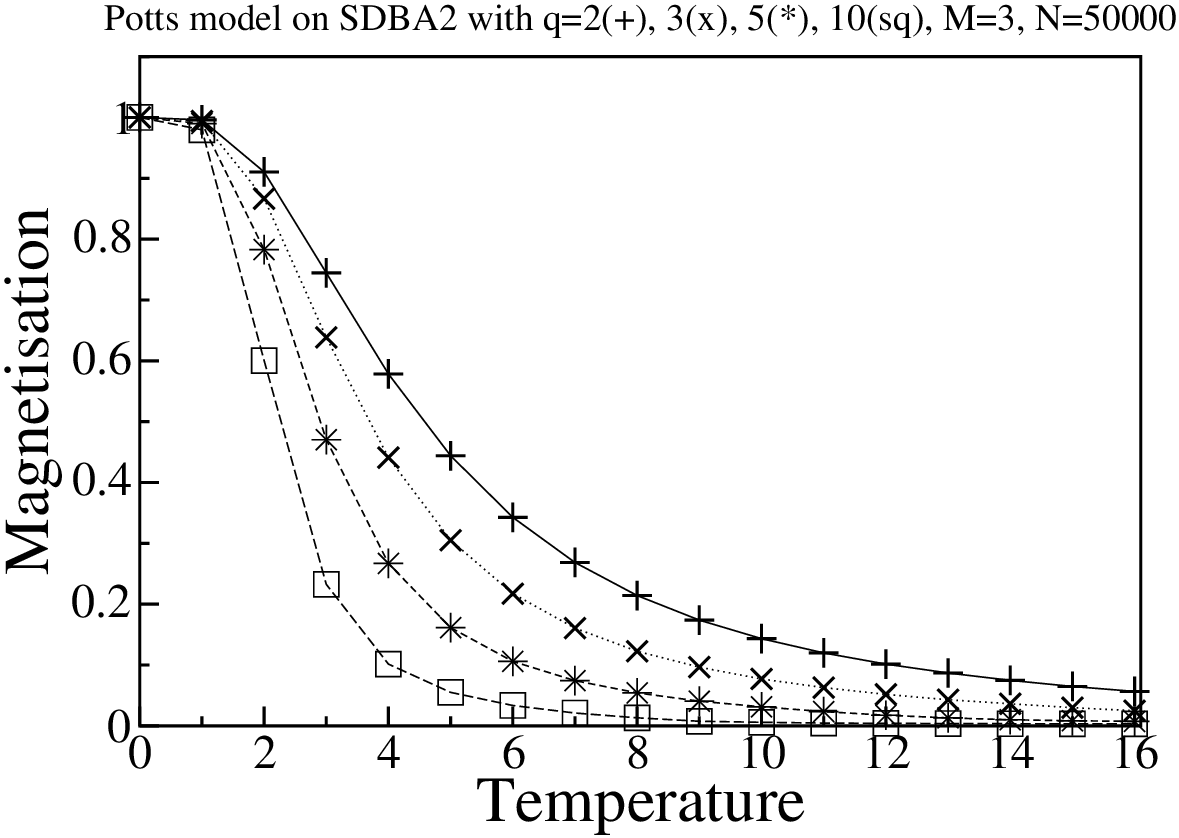}
\end{center}
\caption{As previous figure but now for SDBA2.} 
\label{Fig60}
\end{figure}
\bigskip
\bigskip
\subsection{Potts model}

To study the $q=2$ Potts model we start with all spins 
ordered $S=1$, a number of spins equal to 
$N=500$, $5000$, $50000$, and $100000$ with MCSN = 
$10^{8}$, $10^{7}$ and $2 \times 10^{6}$ 
with HeatBath algorithm, respectively, 
in Figs. \ref{Fig10} (a), (b), (c), and (d).
The temperature is measured in units of $J/k_B$.
We determine the time $\tau$ after which 
the magnetisation has flipped its sign for first time, 
and then take the median value of 9 samples. So this way
is possible to determine various temperature for
different networks size and to extrapolating
$\tau$ to infinity and obtain the critical temperature
for SDBA1 and SDBA2 networks .
\begin{figure}[!ht]
\begin{center}
\includegraphics[angle=0,scale=0.38]{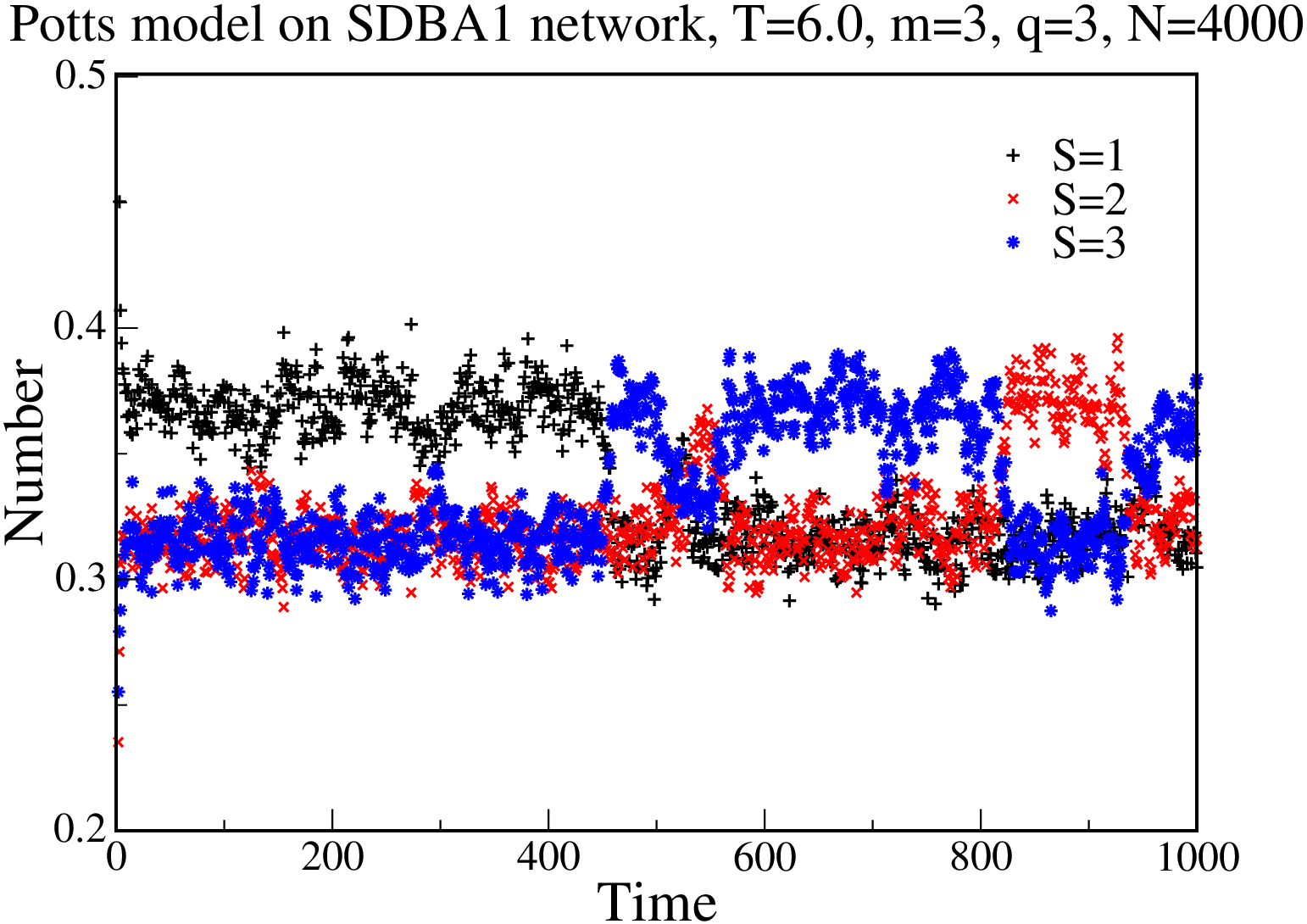}
\end{center}
\caption{Plot of the number of $S=1$, $2$, 
and $3$ states versus the time 
for Potts model with $q=3$ states on SDB1 network, 
with $m=3$, $N=4000$.}
\label{Fig70}
\end{figure}
\bigskip
\begin{figure}[!hb]
\begin{center}
\includegraphics[angle=0,scale=0.38]{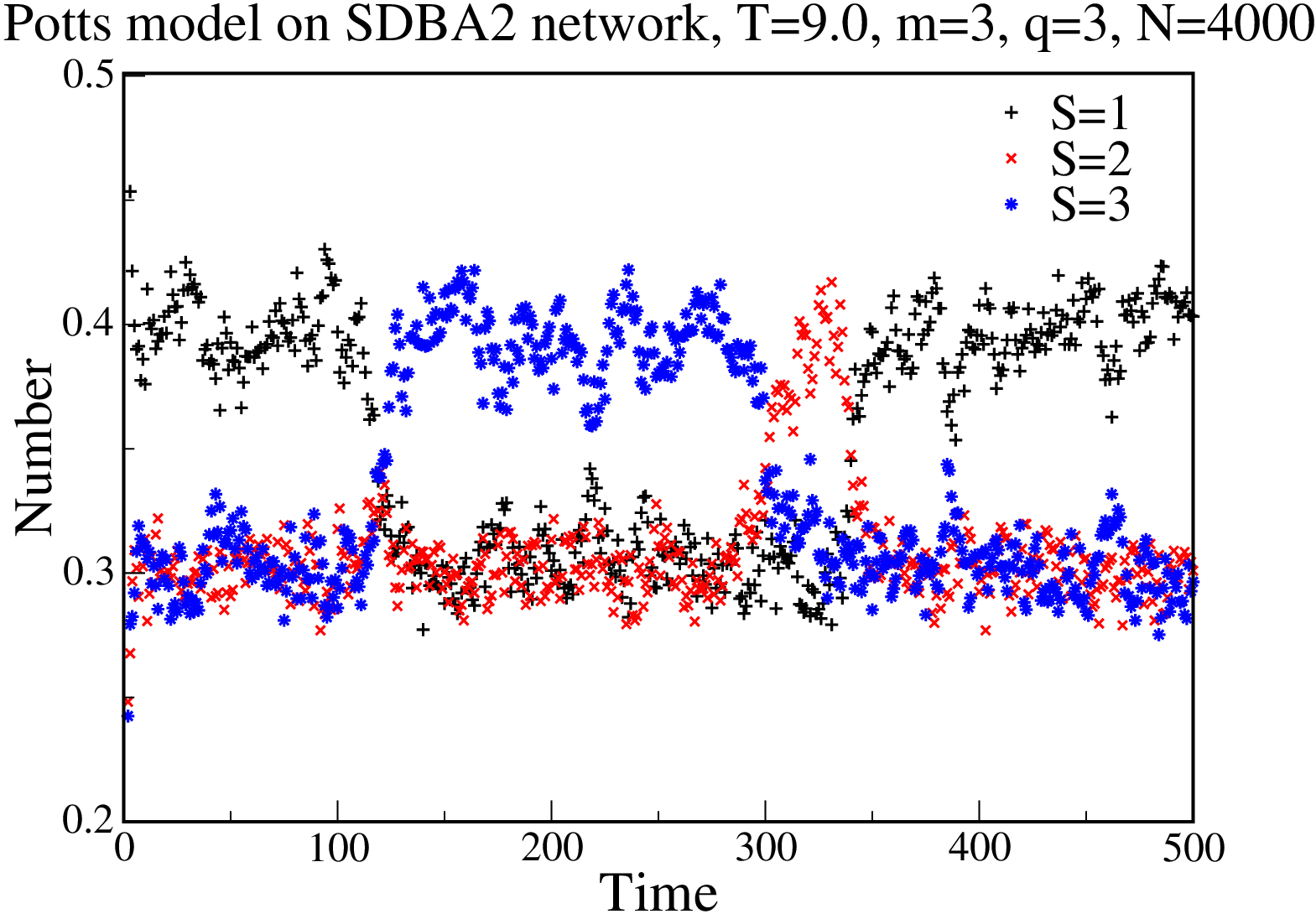}
\end{center}
\caption{The same plot of Fig. \ref{Fig70}, but now for SDBA2 networks. }
\label{Fig80}
\end{figure}
Our simulations on SDBA1 and SDBA2 networks indicate that the $q=2$ Potts
model does not display a usual phase transition and the plots of
the time $1/\ln (\tau)$ versus temperature in Figs. \ref{Fig10}(a), (b), (c), and (d)
show that our results agree with a Vogel-Fulcher(-Tammann) law for the 
relaxation time $\tau$, defined as the first time when the sign of the
magnetisation flips: $1/\ln(\tau)\propto T - T_c(N)$.

We extrapolate $T_c(N)$ for $N=20$, 30, 50, 100, 200, 500, 5000, 50000, and 100000
 to $\approx$ 0.1, 0.2, 0.7, 1.0, 1.1, 1.3, 3.1, 5.3, and 6.0 for SDBA1
and to 0.5, 0.7, 1.0, 1.5, 1.8, 2.5, 4.2, 6.6 and 7.0 for SDBA2, 
increasing roughly logarithmically with $N$
as in \cite{A_H_S}.

In Fig. \ref{Fig40} we show the magnetisation versus 
temperature behaviour on
SDBA1 and SDBA2 network for Potts model with $q=2$ states 
and $m=3$ initial neighbours and 
 $N=50000$ sites. Both SDBA1 and SDBA2 network present 
similar behaviour, but SDBA1
decreases faster than SDBA2 with 
increasing temperature, since SDBA2 has more neighbours than SDBA1.
We use two different programs for Potts and Ising which agreed
 in their results for $q=2$, and they should.

In Fig. \ref{Fig50} we show magnetisation versus temperature on
SDBA1 network for Potts model with $q=2$, 3, and 10 states,
$m=3$ initial neighbours and 
 $N=50000$ sites. Here, we see that increasing  $q$ of Potts model 
provides a more rapid decay of the magnetisation 
as a function of temperature.
In Fig. \ref{Fig60} we show the same behaviour, but now on SDBA2 network. 

In figure \ref{Fig70} we show the time dependence of the number
of $S=1$ and 3 states 
for Potts model with $q=3$ states on SDBA1 network. Here we 
observe the tunneling between these three states with 
the evolution of time.
In Fig. \ref{Fig80} we show the same behaviour as in Fig. \ref{Fig70},
but now on SDBA2 network.
\bigskip
\section{Conclusion}

Finally, for both SDBA1 and SDBA2 networks we found a Vogel-Fulcher law,
suggesting stable ferromagnetism for $T < T_c(N)$. 
For Potts model with $q=3$ on SDBA1 and SDBA2 networks we found a
tunneling between these three states with 
the evolution of time. Similarly to the 
Ising model on undirected Barab\'asi-Albert 
network \cite{A_H_S} there is no usual ferromagnetic transition on these 
SDBA1 and SDBA2 networks, since the curves of magnetisation versus 
temperature are not curved in the usual way at least for $q$ up to $10$, 
and the time dependence in the Ising case ($q=2$) suggests a transition 
temperature increasing logarithmically with increasing system 
size, i.e., $T_c(N)$ increases
roughly logarithmically with network size $N$. Perhaps the limit
$q \rightarrow \infty$ would show a more usual behaviour as Fig.6 and 7 allow.
The distribution of the number of neighbours of each node
decays with a non-universal exponent depending on $m$,
as in \cite{sumuor3}.

The authors are grateful to Dietrich Stauffer for stimulating 
discussions and for a critical reading of the manuscript.
F. W. S. L.  acknowledges the
Brazilian agency FAPEPI (Teresina-Piau\'{\i}-Brasil) 
and CNPQ for  its financial support and. This
work also was supported the system SGI Altix 1350 
the computational park
CENAPAD.UNICAMP-USP, SP-BRAZIL and 
Dietrich Stauffer Computational Physics Lab-TERESINA-PIAU\'I-BRAZIL.

\newpage
\section{Appendix}

This is the Fortran program for Ising model on SDBA2. For SDBA1 a program is
given in \cite{sumuor3}, without spins. 

\noindent

{\small

\begin{verbatim}

       parameter(kb=30000)
C      maxtime=sites
       parameter(nrun=100,maxtime=10000,m=3,iseed=1,max=maxtime+m,
     1 length=1+(1+m)*maxtime+m*(m-1))
       integer*8 ibm,iex,summag,imag
       integer*4 mag
       real*8 factor,ex
       dimension is(max),iex(-kb:kb),neighb(max,kb)
       dimension k(max), nk(kb), list(length)
       data nk/kb*0/,nsteps/100000/,k/max*0/
       print *, max,m,nsteps,nsteps,iseed
       ibm=2*iseed-1
       factor=(0.25d0/2147483648.0d0)/2147483648.0d0
       do 9 itemp=100,1600,+100
       T = 0.01*itemp
       do 5 irun=1,nrun
       do 3 i=1,m
       do 7 j=(i-1)*(m-1)+1,(i-1)*(m-1)+m-1
7      list(j)=i
       jj=0
       do 71 j=1,m
       if(j.eq.i) goto 71
       jj=jj+1
       neighb(i,jj) = j
71     continue
3      k(i)=m-1
       L=m*(m-1)
       if(m.eq.1) then
       L=1
       List(1)=1
       k(1)=1
       neighb(1,1)=1
       endif
C      All m initial sites are connected with each other
       do 1 n=m+1,max
       do 2 new=1,m
4      ibm=ibm*16807
       j=1.d0+(ibm*factor+0.5d0)*L
       if(j.le.0.or.j.gt.L) goto 4
       j=list(j)
       k(n)=k(n)+1
       if(new.eq.1) k(j)=k(j)+1
       if(k(j).gt.kb) stop 9
       list(L+new)=j
c      n selects m sites j which will all influence n
c      n will influence only the first selected j
c      j is always added to LIST, n is added only once
c      k(i) is the number of sites neighb(i, . ) which will influence i
       neighb(n ,new)=j
       if(new.eq.1) neighb(j,k(j))=n
2      continue
       list(L+m+1)=n
       L=L+m+1
1      k(n)=m
       do 5 i=1,max
       k(i)=min0(k(i),kb)
5      nk(k(i))=nk(k(i))+1
C*******************      ISING PART
       DO 20 I=1,MAX
20     IS(I)=1
       DO 21 IE=-KB,KB,1
       EX=EXP(-2*IE/T)
       IF(IE/T.LE.-20.0) EX= 1.0D9
       IF(IE/T.GE. 20.0) EX= 1.0D-9
 21    IEX(IE)=2147483648.0D0*(4.0D0*EX/(1.0D0+EX)-2.0D0)*2147483648.0D0
       SUMMAG=0
       DO 22 MC=1,NSTEPS
       DO 23 I=1,MAX
       IE=0
       DO 24 NB=1,K(i)
24     IE=IE+IS(NEIGHB(I,NB))
       IE=IS(I)*IE
       IBM=IBM*16807
       IF(IBM.LT.IEX(IE)) IS(I)=-IS(I)
23     CONTINUE
       MAG =0
       DO 25 I=1,MAX
25      MAG=MAG+IS(i)
       IMAG=IABS(MAG)
22     IF(MC.GT.(NSTEPS/2)) SUMMAG=SUMMAG+IMAG
       AVERGESUMMAG=SUMMAG*2.0/(MAX*NSTEPS)
C      End of Ising part
9      PRINT *,T,AVERGESUMMAG
       STOP
       END
\end{verbatim}
}


\begin{thebibliography}{99}

\bibitem{me} N. Metropolis, A. W. Rosenbluth, M. N. Rosenbluth, A. H. Teller, 
E. Teller, J. Chem. Phys. {\bf 21}, 1087 (1953).
\bibitem{s-w} R. H. Swendsen, J.-S. Wang, Phys. Rev. Lett. {\bf 58}, 86 (1987).
\bibitem{w-l} F. Wang, D. P. Landau, Phys. Rev. Lett. {\bf 86}, 2050 (2001).
\bibitem{BA}A.-L. Barab\'asi, R. Albert, Science {\bf 286}, 509 (1999).
\bibitem{A_H_S} A. Aleksiejuk, J.A. Ho\l yst, D. Stauffer, Physica A {\bf 310}, 260 (2002) .
\bibitem{sumuor1} M. A. Sumour, M.M. Shabat, Int. J. Mod. Phys. C {\bf 16}, 585 (2005).
\bibitem{sumuor2} M. A. Sumour, M.M. Shabat, D. Stauffer, Islamic University Journal (Gaza) {\bf 14}, 209 (2006). cond-mat/0504460 at www.arXiv.org.
\bibitem{stauffer} D. Stauffer, S. Moss de Oliveira,
P.M.C de Oliveira and J.S.S\'a Martins, {\it Biology, Sociology, Geology by Computational Physicists}.
Elsevier, Amsterdam (2006).
\bibitem{sumuor3} M. A. Sumour and M. A. Radwan, Int. J. Mod. Phys. C {\bf 23},
article 1250062 (2012).
\bibitem{preprint} M. A. Sumour, F. W. S. Lima, M. A. Radwan, and M.M. Shabat, Islamic University Journal (Gaza), to be published.
\end{thebibliography}
\end{document}